\documentclass{jaa}
\usepackage{natbib}
\usepackage{booktabs}
\usepackage{graphicx}

\usepackage{multirow} 

\usepackage{hyperref}

\usepackage{subfig}

\begin{document}\sloppy

\title{\textit{Curvit}:\\ An open-source Python package to generate light curves from UVIT data}


\author{P. Joseph\textsuperscript{1,2,*}, C. S. Stalin\textsuperscript{1}, S. N. Tandon\textsuperscript{3} and S. K. Ghosh\textsuperscript{4}}
\affilOne{\textsuperscript{1}Indian Institute of Astrophysics, Bangalore 560034, India.\\}
\affilTwo{\textsuperscript{2}Department of Physics, CHRIST (Deemed to be University), Bangalore 560029, India.\\}
\affilThree{\textsuperscript{3}Inter-University Centre for Astronomy and Astrophysics, Pune 411007, India.\\}
\affilFour{\textsuperscript{4}Tata Institute of Fundamental Research, Mumbai 400005, India}


\twocolumn[{

\maketitle

\corres{prajwelpj@gmail.com}

\msinfo{1 January 2015}{1 January 2015}

\begin{abstract}

\textit{Curvit} is an open-source Python package that facilitates the creation of 
light curves from the data collected by the Ultra-Violet Imaging Telescope
(UVIT) onboard \textit{AstroSat}, India's first multi-wavelength astronomical 
satellite. The input to \textit{Curvit} is the calibrated events list generated 
by the UVIT-Payload Operation Center (UVIT-POC) and made available
to the principal investigators through the Indian Space Science Data Center. 
The features of \textit{Curvit} include 
(i) automatically detecting sources and generating light curves for all 
the detected sources and 
(ii) custom generation of light curve for any particular source of interest. 
We present here the capabilities of \textit{Curvit} and demonstrate its usability on 
the UVIT observations of the intermediate polar FO Aqr as an example. 
\textit{Curvit} is publicly available on GitHub at \url{https://github.com/prajwel/curvit}.
\end{abstract}


}]

\doinum{12.3456/s78910-011-012-3}
\artcitid{\#\#\#\#}
\volnum{000}
\year{0000}
\pgrange{1--}
\setcounter{page}{1}
\lp{1}

\section{Introduction}
The Ultra Violet Imaging Telescope (UVIT; \citealt{tandon2017CurrentScience}, 
\citealt{Tandon2017performance}), consisting of two co-aligned telescopes of 
aperture 375 mm each, is one of the payloads onboard \textit{AstroSat}. 
\textit{AstroSat} is India's first multi-wavelength astronomical observatory, 
launched by the Indian Space Research Organisation (ISRO) on 28, September 2015
(\citealt{AGRAWAL2006}).
In addition to UVIT, there are three co-aligned X-ray payloads on \textit{AstroSat} 
enabling simultaneous observation of a celestial source over a wide range of 
wavelengths from hard X-rays to the Ultraviolet (UV) band. 
UVIT, with a field of view of 28 arcminute diameter, can perform imaging
and low-resolution slit-less spectroscopy. 
UVIT has a large number of filters and with selectable filters or gratings,
simultaneous observations in far-ultraviolet (FUV, 1300-1800 \AA), 
near-ultraviolet (NUV, 2000-3000 \AA), and visible (VIS, 3200-5500 \AA) channels
are possible. 
Of the three channels, VIS channel is used only for aspect correction,
while the NUV and FUV channels are used for science observations.
The detectors used in all the three channels are intensified CMOS imagers
with 512$\times$512 pixels.
The telescope pointing can drift up to $3'$ over the night time of an orbit
(duration can be up to a maximum of 1800 seconds) with a rate of $3''$/second. 
This drift of the satellite is estimated, as a function of time, using 
observations obtained with the VIS channel. 
In the normal mode of operations, observations are carried out in the full 
window mode (default mode, 512$\times$512 pixels) covering 28 diameter arcminute 
field resulting in a read rate of $\sim$28.7 frames/second.
It is also possible to observe a partial field (that is selectable by the 
principal investigators (PIs) of the observing proposals) read at a higher rate. 
For example, the observation of a small window (100$\times$100 pixels) will provide
640 frames per second. 
The NUV and FUV images are generated by combining short exposure frames with 
shift and add algorithm. 
However, for bright fields, self drift correction of NUV data with 
NUV and FUV data with FUV is also possible.
Two modes of operation exist in UVIT: 
(1) photon counting mode, achievable through high electron multiplication
via high voltage to the microchannel plate of the intensified imager, in the 
case of NUV and FUV channels, where the intensified detectors record the X and Y 
positions of the photons on the detectors and their arrival times and 
(2) integration mode, achievable through a lower electron multiplication,
for VIS channel where the readout consists of image frames 
with a time resolution of 1 second. 
In the final image obtained using photon counting mode, each 
detector pixel is mapped to 8$\times$8 sub-pixels by the centroiding of 
photon events with each sub-pixel having a plate scale of $0.416''$
(\citealt{Hutchings_2007}; \citealt{Postma_2011}).
Due to the availability of time-tagged events in photon counting mode in 
FUV and NUV channels, it is possible to probe the time variability of observed 
sources in UV and generate light curves similar to other branches of 
high energy astronomy such as X-rays and $\gamma$-rays. 
UVIT has been performing as per specifications. 
More details and related calibration can be found in \cite{Tandon2017calibration} 
and \cite{Tandon_2020}.

The UVIT Payload Operation Center (POC) at the Indian Institute of Astrophysics (IIA)
runs the UVIT Level-2 pipeline (UL2P; Ghosh et al. 2021, in preparation) on 
the Level-1 (L1) data, containing both spacecraft and observational data in FITS format, 
to produce science ready Level-2 (L2) data products. 
A description of UVIT data reduction is given in \cite{postma2017ccdlab} 
as well as Ghosh et al. (2021).
The UVIT-POC processed L1 and L2 data are made available to the PIs of the observations
through the ISRO Indian Space Science Data Center (ISSDC).
The L2 data contains 
(i) orbit-wise calibrated events list after corrections for the drift of the spacecraft, 
flat field, and distortion 
(ii) orbit-wise science ready images in detector coordinate and world 
coordinate systems and 
(iii) combined images that belong to a single pointing, wherein observations in a 
particular filter carried out over many orbits in a particular pointing are combined.
L2 data from ISSDC are science ready products, and the PIs can directly carry 
out their photometric, spectroscopic or imaging analysis. 
Alternatively, PIs willing to do a custom analysis of their observations can
also do so, using UL2P, along with the CALDB (that contains the 
calibration data files) and the CATALOG (that contains the catalogues for astrometry)
downloadable from ISSDC\footnote{http://www.issdc.gov.in/}, 
UVIT-POC\footnote{http://uvit.iiap.res.in/} or 
the AstroSat Science Support cell\footnote{http://astrosat-ssc.iucaa.in/}. 

Time variable phenomenon can be studied naturally using the "photon counting mode" 
of operation for the UV bands of image acquisition by UVIT.
In principle, in the full window mode observations with UVIT, one can study 
time-varying phenomenon with a time resolution as low as 66 msec in both the 
UV bands. 
Even higher time resolution is possible with smaller window observations 
(for example, $\sim$3 msec in 100$\times$100 pixels window).
Studies of such high-resolution events will open up a new avenue of research in 
UV Astronomy, and for such studies, software tools are required to generate 
the light curves directly from the events list. 
The motivation is therefore to develop a software tool, that has the ability 
to create light curves from the events list.
Here we present \textit{Curvit}, an open-source Python package designed 
to create light curves from UVIT L2 events list. 
\textit{Curvit} makes use of the functionalities available in other open-source Python 
packages such as Astropy (\citealt{astropy:2013}; \citealt{astropy:2018}), 
NumPy (\citealt{numpy}), Matplotlib (\citealt{matplotlib}), 
Photutils (\citealt{photutils}, and Scipy (\citealt{scipy}). 
The availability of this tool to the PIs of UVIT will also avert the cumbersome task
of first creating images of small time-bins and then doing the photometry of the 
target to generate light curves. 
Section 2 contains a short summary of the L2 products at ISSDC. 
In Sections 3 and 4, we describe the functionalities and working of the tool.
In the final section, we demonstrate the usefulness of the tool by 
generating the light curve of intermediate polar FO Aquarii (FO Aqr), 
observed under the guaranteed time and now open for public after 
the lock-in-period. 

\section{UVIT data products at ISSDC}
The UVIT data is available at ISSDC AstroBrowse website\footnote{https://astrobrowse.issdc.gov.in/astro\_archive/archive/Home.jsp}. 
Both L1 and L2 data of UVIT are available as compressed files at the archive. 
L2 products are organised into two categories: 
individual datasets (single orbit for a filter and window; see Table \ref{singleorbit}), 
and combined datasets over all the orbits (for a single filter and window; see Table \ref{combinedorbit}).
The UVIT filters are given in Table \ref{filters}. 

\begin{table*}
\caption{Details of the data products sent to ISSDC for each orbit of observation done in a particular window-size and filter.}
\begin{tabular}{|c|c|c|c|c|c|c|}
\hline
\multirow{2}{*}{Product}                                                              & \multirow{2}{*}{Description}                                    & \multicolumn{2}{c|}{RAS VIS$^{\ast}$} & \multicolumn{2}{c|}{RAS NUV$^{\ast}$} & \multirow{2}{*}{Total} \\ \cline{3-6}
                                                                                      &                                                                 & NUV           & FUV           & NUV           & FUV           &                        \\ \hline
\begin{tabular}[c]{@{}c@{}}Sky Image \\ (Instrument \\ coordinates)\end{tabular}      & \begin{tabular}[c]{@{}c@{}}4800$\times$4800 sub-pixel$^2$\\FITS image\end{tabular} & 1             & 1             & 1             & 1             & 4                      \\ \hline
\begin{tabular}[c]{@{}c@{}}Sky Image \\ (Astronomical \\ coordinates)\end{tabular}    & \begin{tabular}[c]{@{}c@{}}4800$\times$4800 sub-pixel$^2$\\ FITS image\end{tabular} & 1             & 1             & 1             & 1             & 4                      \\ \hline
\begin{tabular}[c]{@{}c@{}}Exposure Map \\ (Astronomical \\ coordinates)\end{tabular} & \begin{tabular}[c]{@{}c@{}}4800$\times$4800 sub-pixel$^2$\\ FITS image\end{tabular} & 1             & 1             & 1             & 1             & 4                      \\ \hline
\begin{tabular}[c]{@{}c@{}}Error Map \\ (Astronomical \\ coordinates)\end{tabular}    & \begin{tabular}[c]{@{}c@{}}4800$\times$4800 sub-pixel$^2$\\ FITS image\end{tabular} & 1             & 1             & 1             & 1             & 4                      \\ \hline
\begin{tabular}[c]{@{}c@{}}Photon Events \\ List\end{tabular}                         & \begin{tabular}[c]{@{}c@{}}FITS binary \\ table\end{tabular}    & 1             & 1             & 1             & 1             & 4                      \\ \hline
RAS file                                                                              & \begin{tabular}[c]{@{}c@{}}FITS binary \\ table\end{tabular}    & \multicolumn{2}{c|}{1}        & \multicolumn{2}{c|}{1}        & 2                      \\ \hline
\end{tabular}

\vspace{0.5em}

\noindent $^{\ast}$The above file structure corresponds to the ideal case when VIS, NUV, and FUV are configured by the PI. In the event of VIS not being configured, the files generated using the relative aspect series (RAS) obtained from VIS data will be missing. Similar is the case for NUV.
\label{singleorbit}
\end{table*}

\begin{table*}
\caption{Details of the combined data products sent to ISSDC. The observations carried out over the entire pointing are combined filter wise.}
\begin{tabular}{|c|c|c|c|c|c|c|}
\hline
\multirow{2}{*}{Product}                                                                & \multirow{2}{*}{Description}                                    & \multicolumn{2}{c|}{RAS VIS} & \multicolumn{2}{c|}{RAS NUV} & \multirow{2}{*}{Total} \\ \cline{3-6}
                                                                                        &                                                                 & NUV            & FUV            & NUV            & FUV            &                        \\ \hline
\begin{tabular}[c]{@{}c@{}}Sky Image -- A$^{\ast}$ \\ (Astronomical \\ coordinates)\end{tabular}   & \begin{tabular}[c]{@{}c@{}}4800$\times$4800 sub-pixel$^2$\\ FITS image\end{tabular} & 1              & 1              & 1              & 1              & 4                      \\ \hline
\begin{tabular}[c]{@{}c@{}}Sky Image -- B$^{\$}$ \\ (Astronomical \\ coordinates)\end{tabular} & \begin{tabular}[c]{@{}c@{}}4800$\times$4800 sub-pixel$^2$\\ FITS image\end{tabular} & 1              & 1              & 1              & 1              & 4                      \\ \hline
\begin{tabular}[c]{@{}c@{}}Exposure Map$^{\dagger}$ \\ (Astronomical \\ coordinates)\end{tabular} & \begin{tabular}[c]{@{}c@{}}4800$\times$4800 sub-pixel$^2$\\ FITS image\end{tabular} & 1              & 1              & 1              & 1              & 4                      \\ \hline
\begin{tabular}[c]{@{}c@{}}Error Map$^{\dagger}$ \\ (Astronomical \\ coordinates)\end{tabular}   & \begin{tabular}[c]{@{}c@{}}4800$\times$4800 sub-pixel$^2$\\ FITS image\end{tabular} & 1              & 1              & 1              & 1              & 4                      \\ \hline
\end{tabular}

\vspace{0.5em}

\noindent $^{\ast}$Sky image -- A: The astrometric accuracy of this image is 
limited to the accuracy of knowledge of the spacecraft aspect, which is 
typically around 2-3 arcmin.

\vspace{0.5em}

\noindent $^{\$}$Sky image -- B: This is the final image generated after 
astrometry which may or may not be successful. 
When the astrometry is successful, the accuracy in aspect is typically 3 arcsec. When the astrometry is not successful, this image is a copy of Sky image -- A. The information about astrometry being successful or unsuccessful is available in the header of the FITS images.

\vspace{0.5em}

\noindent $^{\dagger}$They correspond to the co-ordinate system of Sky image -- B.
\label{combinedorbit}
\end{table*}

\begin{table*}
\caption{The UVIT filters in VIS, NUV and FUV channels.}
\centering
\begin{tabular}{|c|c|c|c|c|c|c|c|c|}
\hline
\multicolumn{3}{|c|}{VIS}                                                                                                         & \multicolumn{3}{c|}{NUV}                                                                                                          & \multicolumn{3}{c|}{FUV}                                                                                                          \\ \hline
Filter ID & \begin{tabular}[c]{@{}c@{}}Old filter\\ name\end{tabular} & \begin{tabular}[c]{@{}c@{}}New filter\\ name\end{tabular} & Filter ID & \begin{tabular}[c]{@{}c@{}}Old filter\\ name\end{tabular} & \begin{tabular}[c]{@{}c@{}}New filter\\ name\end{tabular} & Filter ID & \begin{tabular}[c]{@{}c@{}}Old filter\\ name\end{tabular} & \begin{tabular}[c]{@{}c@{}}New filter\\ name\end{tabular} \\ \hline
F1        & VIS3                                                      & V461W                                                     & F1        & Silica - 1                                                & N242W                                                     & F1        & CaF2 - 1                                                  & F148W                                                     \\ \hline
F2        & VIS2                                                      & V391M                                                     & F2        & NUVB15                                                    & N219M                                                     & F2        & BaF2                                                      & F154W                                                     \\ \hline
F3        & VIS1                                                      & V347M                                                     & F3        & NUVB13                                                    & N245M                                                     & F3        & Sapphire                                                  & F169M                                                     \\ \hline
F4        & ND1                                                       & V435ND                                                    & F4        & Grating                                                   &                                                           & F4        & Grating - 1                                               &                                                           \\ \hline
F5        & BK7                                                       & V420W                                                     & F5        & NUVB4                                                     & N263M                                                     & F5        & Silica                                                    & F172M                                                     \\ \hline
          &                                                           &                                                           & F6        & NUVN2                                                     & N279N                                                     & F6        & Grating - 2                                               &                                                           \\ \hline
          &                                                           &                                                           & F7        & Silica - 2                                                & N242Wa                                                    & F7        & CaF2 - 2                                                  & F148Wa                                                    \\ \hline
\end{tabular}
\label{filters}
\end{table*}

\section{\textit{Curvit} Workflow}
\textit{Curvit}\footnote{https://github.com/prajwel/curvit} is an open-source Python package
to produce light curves from UVIT data. 
The events list from the official UVIT L2 pipeline (version 6.3 onwards) is 
required as an input to the package. 
\textit{Curvit} has two functions for light curve creation; \texttt{makecurves} and \texttt{curve}. 
Both the functions accept a single events list at a time which the user has to provide.
We describe below each of these functions and its usage on 
the observation of the intermediate polar FO Aqr.

\subsection{\texttt{makecurves}}
The \texttt{makecurves} function of \textit{Curvit} automatically detects sources
from the events list and create light curves for all of them. 
The user will have a control on the number of sources detected 
automatically through the use of the detection threshold parameter. 
Two source detection methods are available; '\textit{daofind}' and '\textit{kdtree}'.
The user can select the preferred source detection method using the
$detection\_method$ parameter (the default value is '\textit{daofind}').

The '\textit{daofind}' method detects sources in the following manner.
It first creates a 4800$\times$4800 sub-pixel$^2$ image from the events list and 
a circular mask is applied to select the central $\sim$24 arcminute region.
Sources are then detected using the \textit{daofind} algorithm (\citealt{stetson1987daophot}).
Mean and standard deviation values of the background, required by \textit{daofind},
are estimated by $Curvit$ itself and the user can control the number of sources 
detected using the $threshold$ parameter. 
Pixels in the image that have the events greater than the threshold times 
the standard deviation of the background will be detected.

The '\textit{kdtree}' method works as follows. 
A source is characterised by a cloud of events around its centroid. 
Therefore, to detect sources, the events are projected onto a two-dimensional 
Cartesian grid (with one grid cell being of 1 sub-pixel$^2$ size), and the
grid cells are sorted based on the number of events falling in each cell. 
Since a single source can occupy multiple grid cells, a nearest neighbour 
search using \textit{kdtree} algorithm is performed to remove grid cells 
belonging to the same source (\citealt{maneewongvatana1999s}).
This method may not work properly on crowded fields.
But in non-crowded fields, the method can detect all the sources present 
in the events list. 
However, the user may limit the number of sources to be detected 
using the parameter $how\_many$.
Also, the aperture radius that is used to count the source events 
(through $radius$) and the size of the time bin to generate the light curves 
(using the parameter $bwidth$) can be controlled by the user.

\begin{figure*}
\hspace*{-0.2cm}\includegraphics[scale=0.8]{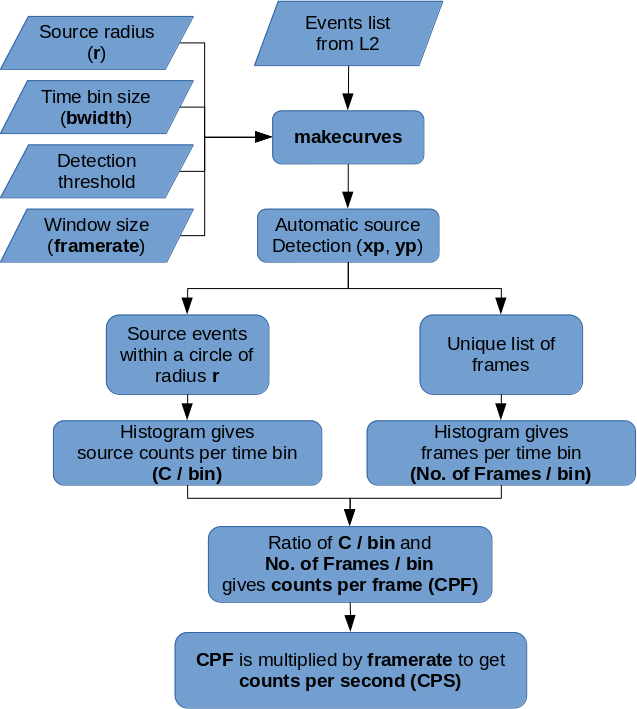}
\caption{Flowchart for the function \texttt{makecurves} in \textit{Curvit}.}\label{makecurvesflow}
\end{figure*}

The function \texttt{makecurves} generates light curves for each 
detected sources that depend on the threshold set by the user in terms 
of decreasing order of brightness. 
The operation of the function is summarised in Fig. \ref{makecurvesflow}.

\subsection{\texttt{curve}}
\begin{figure*}
\hspace*{-0.2cm}\includegraphics[scale=0.8]{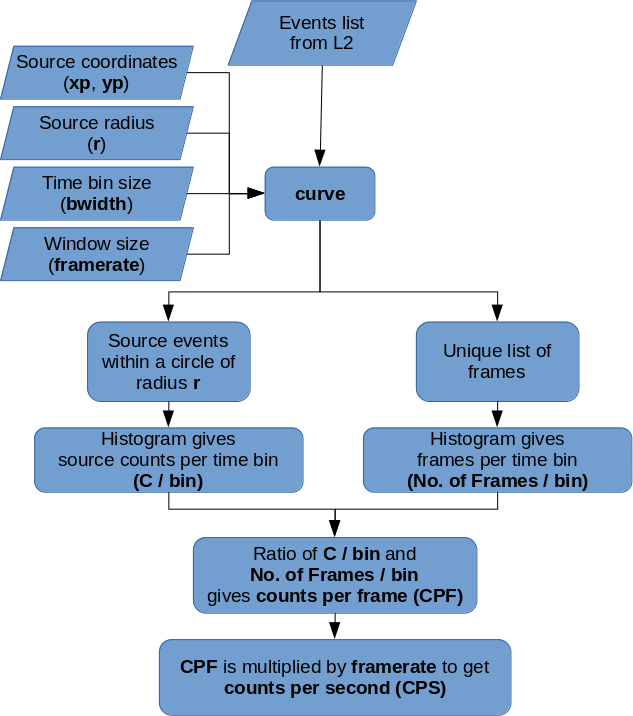}
\caption{Flowchart for the function \texttt{curve} in \textit{Curvit}.}\label{curveflow}
\end{figure*}

The function \texttt{curve} is similar to \texttt{makecurves}, with 
the exception that it generates light curve for a single user-defined source 
(through $xp$ and $yp$ parameters).   
Here also, the user can specify the source aperture radius and the binning time. 
Its operation is summarised in the Fig. \ref{curveflow}.

\section{Light curve Creation}

From the events list FITS table, the columns 'Fx', 'Fy', 'MJD\_L2', and 
'EFFECTIVE\_NUM\_PHOTONS' (hereafter ENP) are used by \textit{Curvit} 
(see Table \ref{eventslist}).
However, the events list FITS table has more columns than the ones 
given in Table \ref{eventslist}.
Each row of the table characterises a single event defined by X and Y Cartesian 
coordinate positions ('Fx' and 'Fy') with an associated time value ('MJD\_L2'). 
For the L2 data available from ISSDC, 'MJD\_L2' provides 
only an approximate absolute time, good to $\sim$1 second.
The time values in 'MJD\_L2' column increment as the frame number (as denoted in 
the 'FrameCounts' column of events list) changes. 
Therefore, 'MJD\_L2' column can be considered as a proxy for 'FrameCounts' column. 
ENP column stores the 'counts/second' contribution from that specific event 
(row of the table) after including instrumental correction like flat-field 
across the detector (Ghosh et al. 2021, in preparation). 
In the methods mentioned below, each event is weighted as per 
the corresponding ENP value.

\begin{table}
\caption{Sample events list that show only the columns used by Curvit.}
\centering
\begin{tabular}{@{}ccccc@{}}
\toprule
Frame\\Counts & Fx     & Fy     & ENP  & MJD\_L2       \\ \midrule
…          & …      & …      & …    & …             \\
3          & 2461.9 & 2918.0 & 28.5 & 213453019.048 \\
3          & 3139.5 & 3651.2 & 25.6 & 213453019.048 \\
4          & 875.9  & 2924.2 & 23.8 & 213453019.084 \\
4          & 1444.0 & 3605.5 & 23.7 & 213453019.084 \\
4          & 2166.7 & 3934.6 & 24.8 & 213453019.084 \\
5          & 3355.4 & 1229.8 & 25.6 & 213453019.120 \\
5          & 3216.9 & 1497.7 & 26.3 & 213453019.120 \\
5          & 2798.5 & 3836.8 & 25.7 & 213453019.120 \\
6          & 3113.4 & 2230.7 & 27.5 & 213453019.156 \\
6          & 4367.8 & 2483.6 & 20.7 & 213453019.156 \\
…          & …      & …      & …    & …             \\ \bottomrule
\end{tabular}
\label{eventslist}
\end{table}

For a given source coordinate position, it is possible to define an aperture of some 
radius in the detector coordinate system (x,y) and select only those events 
(rows of the table) which fall inside the aperture. 
Thus, a subset table can be created from the original table. 
Here, the term original table is used to refer to the input events list table.
The subset table will have only those events of the original table,
that is assumed to be solely coming from the source.
We will refer to the time values (equivalent to frame numbers) in 
the subset table as subset-time. 
Also, a separate array of unique time values is created from the original table. 
We will call this array as original-time.

Using the resultant two arrays of time (subset-time and original-time), 
two separate histograms are created. 
The number of bins (same for both histograms) to be used is determined from 
the user-provided bin width (see Appendix A). 
For subset-time, the histogram can be interpreted as the number 
of events per bin coming from the source. 
Whereas histogram of original-time should ideally give a constant value per bin. 
For example, if bin width is set to 1 second, one should always get $\sim$28.7 
frames per second at all the bins for 512$\times$512 mode. 
However, if some frames are missing (for example, due to large telescope drift or 
missing data), this will be reflected as reduced values 
in the original-time histogram. 
Therefore correction for missing frames is carried out using the original-time
histogram (see Appendix B for a detailed explanation).

By taking the ratio of two histograms, counts per frame (CPF) 
array is obtained. 
CPF array is then multiplied by frame-rate ($\sim$28.7 for 512$\times$512 mode;
the user may specify the frame-rate using the $framecount\_per\_sec$ parameter) 
to obtain the counts per seconds (CPS) array. 
Finally, the CPS array is plotted against time as the light curve. 
The user is advised to look for variability in sources within the central 
20 arcminute region to reduce telescope drift effects.

\subsection{Background, aperture, and saturation corrections}
Estimation of background CPS can be obtained by manually specifying 
a background region ($x\_bg$ and $y\_bg$) and aperture size ($sky\_radius$).
It is also possible in \textit{Curvit} to automatically determine background count-rate.
To do this, a two-dimensional (2D) histogram of events with  
16$\times$16 sub-pixel$^2$ bin size is created.
A mask is applied on the 2D histogram to select only the central $\sim$24 arcminute region.
As opposed to a source, a background region will not have the crowding of events 
around some centroid and have a comparatively low number of events in a 
2D histogram bin containing it. 
Also, the values of histogram bins containing background regions are 
assumed to be normally distributed.
Therefore, the bins with sources are removed using sigma clipping of the 
2D histogram values and locations of background histogram bins
are randomly selected to estimate the mean sigma clipped background count-rate 
using an aperture size of $sky\_radius$.

The end-user has also the option to apply aperture and saturation corrections
using $aperture\_correction$ and $saturation\_correction$ parameters.
Depending on the size of the radius used, the measured CPF will change.
This difference is represented as a table of encircled energy at radii 
from 1.5 - 95 sub-pixels for both UV channels in \cite{Tandon_2020}. 
Cubic interpolation was used to create a continuous function mapping 
radius to encircled energy in the stipulated range. 
From the encircled energy, aperture-corrected CPF can be represented 
as follows,

\vspace{-1em}

\begin{equation}
 \mbox{Aperture-corrected CPF} = \frac{\mbox{Measured CPF} \times 100}{\mbox{Encircled energy (\%)}}
\end{equation}

Thus, aperture correction is applied to CPF values in each bin. 

If the average photon rate per frame is not $\ll$ 1, the effects of saturation 
will make the measured CPF different from the real CPF by a value of RCORR.  
As long as the measured CPF is $<$ 0.6, RCORR 
can be estimated as below \citep{Tandon2017calibration},

\vspace{-1em}

\begin{equation}
 \mbox{ICPF5} = -\ln(1 - \mbox{Measured CPF})
\end{equation}

\vspace{-2em}

\begin{equation}
 \mbox{ICORR} = \mbox{ICPF5} - \mbox{Measured CPF}
\end{equation}

\vspace{-2em}

\begin{equation}
 \mbox{RCORR} = \mbox{ICORR} \times (0.89 - (0.30 \times \mbox{ICORR}^2))
\end{equation}

\vspace{-2em}

\begin{equation}
 \mbox{Real CPF} = \mbox{Measured CPF} + \mbox{RCORR}
\end{equation}

Following the method above, saturation correction is applied 
to CPF values in each bin.

\subsection{Zero event frames and centroid parity errors}

When the total count-rate for the observed region is small, then
there will be many frames with no events;
as the total count-rate go down, the fraction of zero event frames go up. 
This can be modelled using Poisson statistics \cite{Tandon2017calibration}.

\begin{figure}
\hspace*{-0.2cm}\includegraphics[scale=0.5]{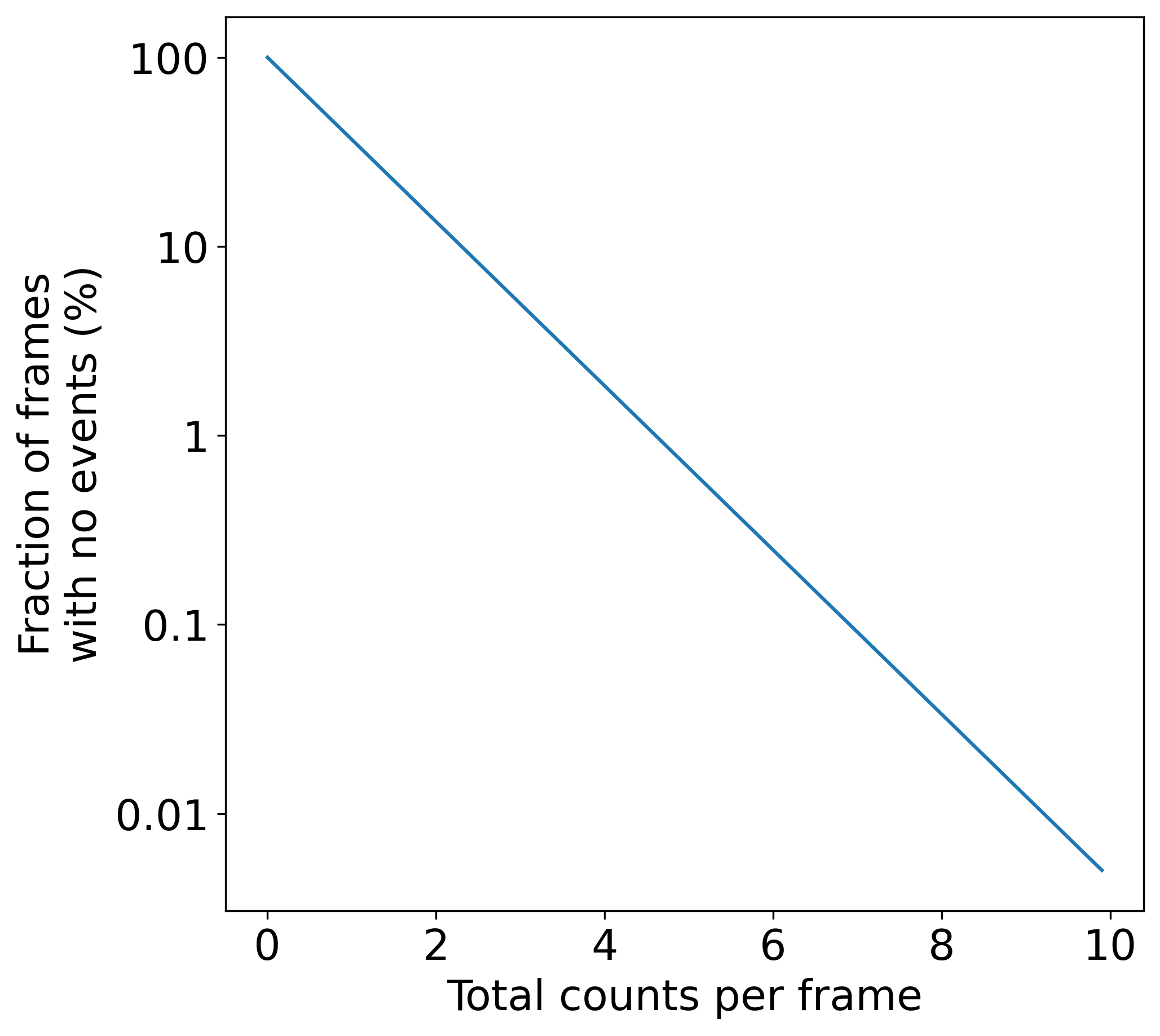}
\caption{Fraction of frames with no events as a function of total counts per frame.}\label{zero_events}
\end{figure}

\vspace{-1em}

\begin{equation}
 F0_{total} = \exp{(-X_{total})}
\end{equation}

\noindent where $X_{total}$ is the CPF for the whole field of view and $F0_{total}$ 
is the fraction of frames with no events (see Fig. \ref{zero_events}).
For $X_{total}$ values above 4.6, the $F0_{total}$ is less than 1\% and 
$F0_{total}$ is less than 5\% for $X_{total}$ values above 3.
Since original-time histogram is used to account for the missing frames, 
CPF values for the source will be overestimated depending on $X_{total}$ value. 

Additionally, a small fraction of the events (less than 0.01\%) 
can be lost due to centroid parity errors. 
Since both zero event frames and centroid parity errors are randomly distributed, 
any light curves with periodicity and/or high count-rate can be taken to be 
having true variability. 

\section{Test on sample data}
\begin{figure}
\centering\includegraphics[width=1.0\columnwidth]{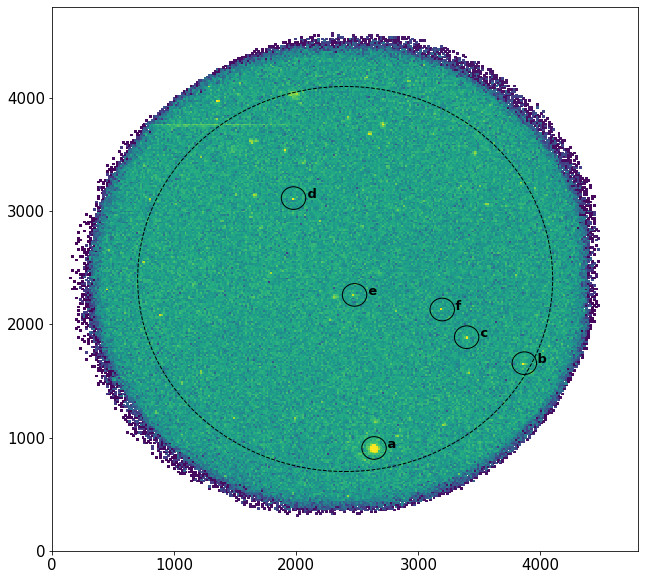}
\caption{The sources that are detected within the central 24 arcminute region (dashed circle) by \texttt{makecurves}.}\label{detections}
\end{figure}

\begin{figure*}
\begin{tabular}{cc}
\subfloat[]{\includegraphics[width = 3.25in]{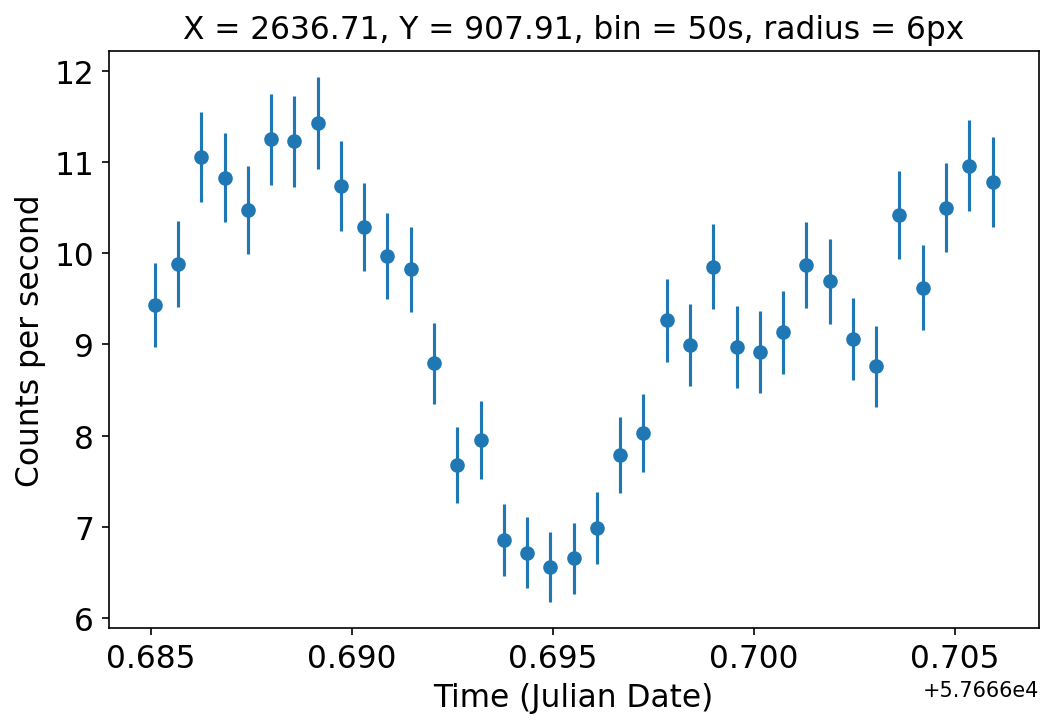}} &
\subfloat[]{\includegraphics[width = 3.25in]{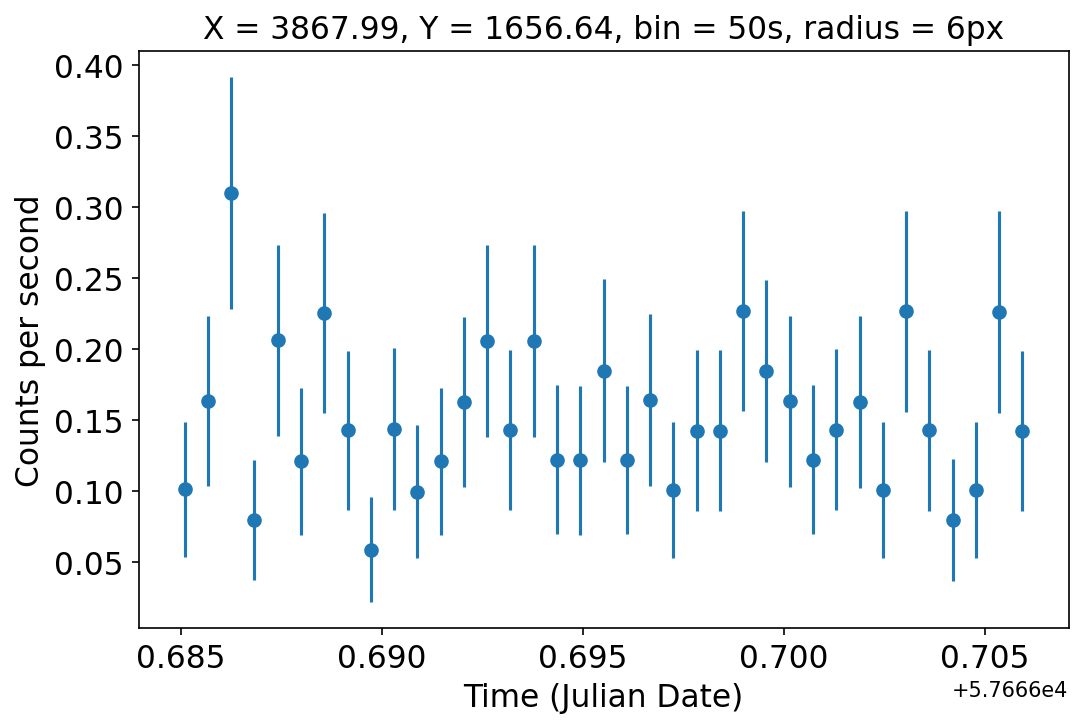}}\\
\subfloat[]{\includegraphics[width = 3.25in]{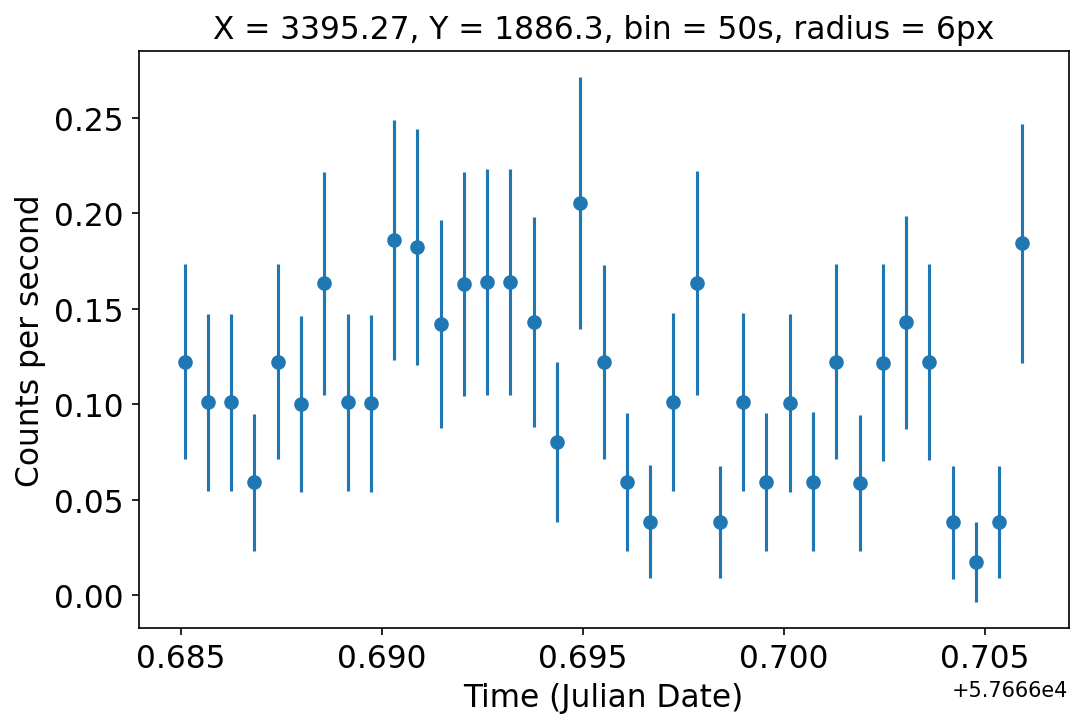}} &
\subfloat[]{\includegraphics[width = 3.25in]{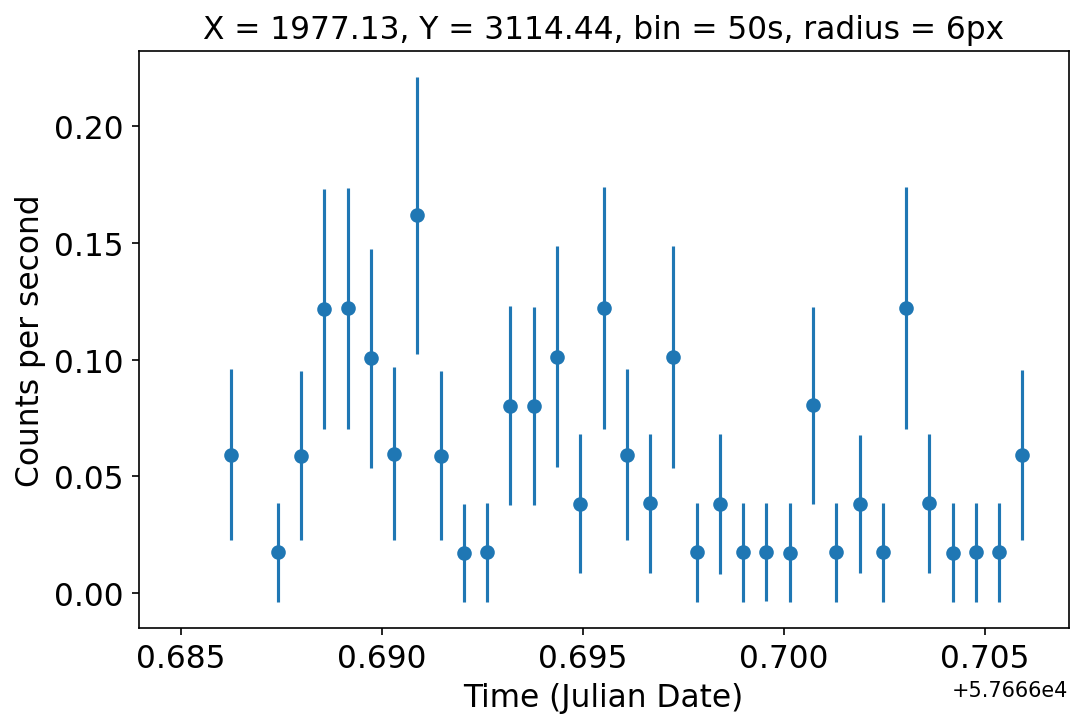}}\\
\subfloat[]{\includegraphics[width = 3.25in]{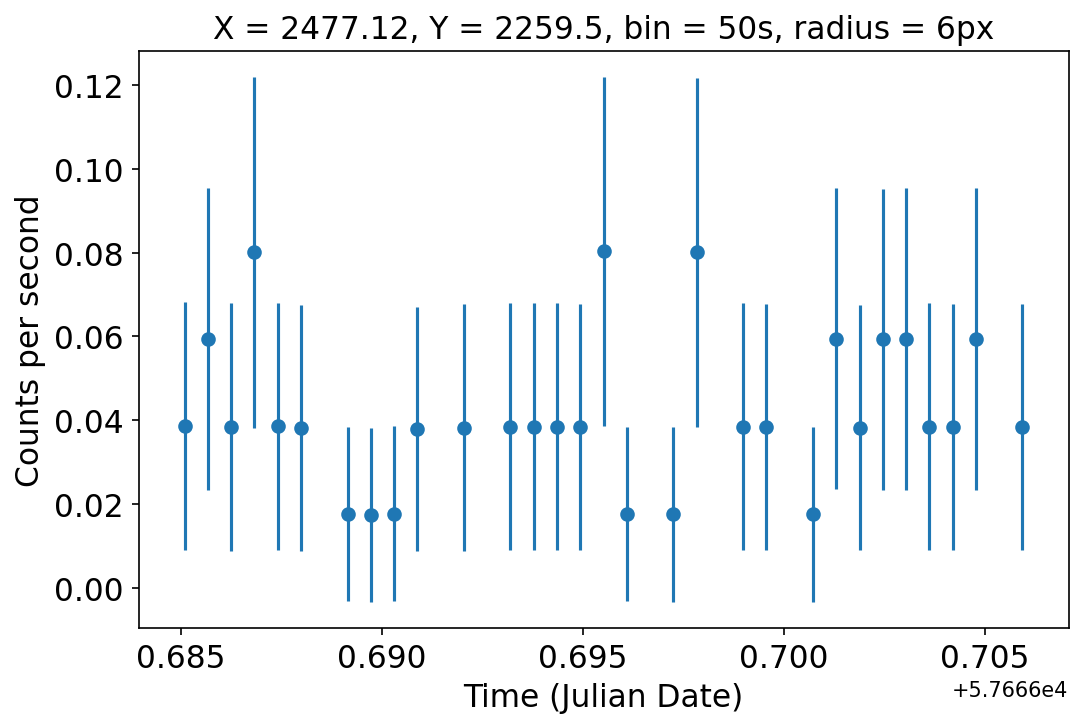}} &
\subfloat[]{\includegraphics[width = 3.25in]{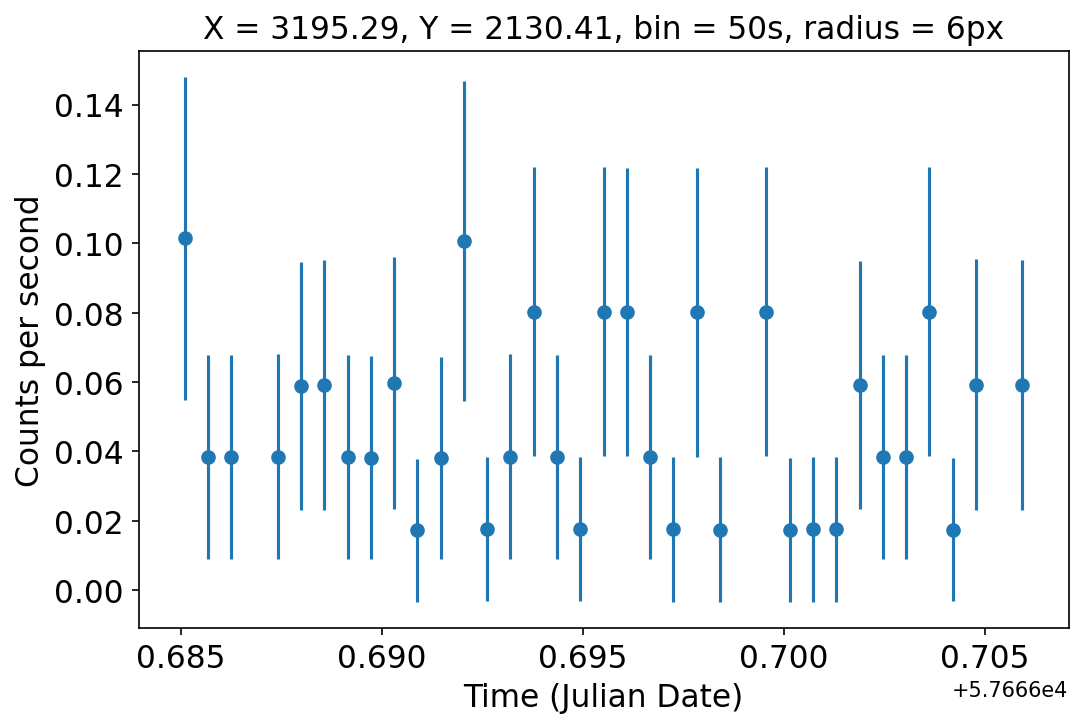}}
\end{tabular}
\caption{Light curves generated by the \texttt{makecurves} function in \textit{Curvit}.}\label{curves}
\end{figure*}

We took the publicly available L2 data (UL2P version 6.3) of FO Aqr 
(observation ID: G06\_084T01\_9000000710) from the ISSDC AstroBrowse website. 
As an example, the \textit{Curvit} software was run on L2 events lists that 
correspond to observations with the FUV filter F148W to create light curves. 
Using \texttt{makecurves} function with the detection threshold value of 4, 
six sources were automatically detected and \textit{Curvit} generated 
light curves for all of them (Fig. \ref{curves}).
they are labelled as (a) to (f) in Fig. \ref{detections}. 
An aperture radius of 6 sub-pixels (2.5 arcseconds) and time bin 
of 50 seconds was used.
The observed variance ($\Sigma^2$) of each of the light curves in 
Fig. \ref{curves} is due to contributions from intrinsic source 
variability and measurement uncertainty. 
In the event of the source being non-variable, the contribution of 
source variability to the observed variance is zero. 
Therefore variance becomes equal to the average value of 
squared errors ($\sigma^2$) of the light curve points. 
For each light curve, we calculated the variance and the mean of 
squared errors in the points and the ratio R =  $\Sigma^2 / \sigma^2$. 
For non-variable sources, this ratio R will be close to unity, and a 
source is considered variable if R is much larger than unity. 
The values of R calculated for the light curves of all the sources 
is given in Table \ref{variability}. 

We also calculated the normalised excess variance, $F_{var}$, 
for the light curves to test the significance of their variations following 
\cite{rani2019study}. 
For variable sources, $F_{var}$ will have a real and positive value 
(see Table \ref{variability}). 
It is evident from Table \ref{variability} that only for the source (a), 
namely FO Aqr, 
(i) the ratio R is much larger than unity and 
(ii) $F_{var}$ is much larger than the error in $F_{var}$. 
These indicate that the larger variance of the source (a) is due 
to the intrinsic variability of the source. 
Though sources (c) and (d) have real and positive $F_{var}$ values, 
and R greater than unity, for (c) the error in $F_{var}$ is much larger 
than $F_{var}$ itself and for (d) $F_{var}$ is not that significant 
considering its error. 
Thus from the light curves of sources (a) to (f), statistical tests confirm 
that only source (a) is variable. 
We note that the light curves of the sources presented here pertain to 
one orbit of data and such an analysis of one orbit data can only pick out 
short-period variables and will miss out long-period variables. 
Therefore, for sources where the period of variability is much longer than 
the one orbit data presented here, it is advisable to generate light curves 
for the complete observation (that can spread over many orbits), which might 
amount to carrying out photometry on each orbit wise images and then check 
for the presence of variability.

\begin{table}
\caption{Variability measure of detected sources.}
\centering
\begin{tabular}{@{}cccc@{}}
\hline
Source & $\Sigma^2 / \sigma^2$ & $F_{var}$ & $F_{var}$ error \\ \midrule
a      & 11.87                 & 0.14    & 0.02          \\
b      & 0.88                  & -       & -             \\
c      & 1.01                  & 0.04    & 0.58          \\
d      & 1.39                  & 0.38    & 0.16          \\
e      & 0.35                  & -       & -             \\
f      & 0.61                  & -       & -             \\ \bottomrule
\end{tabular}
\label{variability}
\end{table}

\section{Conclusion}
The UVIT-POC at IIA processes the L1 data for UVIT received from ISSDC, 
generates science ready L2 products and transfers both the corrected 
L1 and L2 products to ISSDC for archival and dissemination to the PIs. 
Among the various files sent to ISSDC is the calibrated orbit-wise photon events list. 
The \textit{Curvit} software tool presented here is a standalone package in Python 
that can generate light curves from the events list. 
This overcomes the cumbersome task of first creating images at any time resolution 
from the events list and then doing photometry on the images to generate the 
light curves of any desired object in the observed field.  
The \textit{Curvit} package has the capability to 
(1) generate light curves for all the sources in the observed field 
detected above a threshold for any time binning given by the user and 
(2) generate light curve for any particular source at any time binning 
desired by the user. 
We have also shown an example light curve of a target FO Aqr observed by UVIT. 
\textit{Curvit} is publicly available on GitHub at \url{https://github.com/prajwel/curvit}.

\appendix

\section{To estimate the number of bins}
To calculate the number of bins, the very first and last values of 
the original-time array is taken to estimate the width of the time array. 
Then, the time array width is divided by the bin width, and integer part 
of the resultant value is taken as the number of bins.

\begin{equation}
 \mbox{Number of bins} = \frac{\mbox{original time\ width}}{\mbox{bin width}}
\end{equation}

\section{Missing frame correction}
Assume that an ideal non-variable source has a flux of 
0.1 CPF (or $\sim$2.87 CPS for 512$\times$512 mode). 
If the bin width were set to 1 second, then 
the subset-time histogram would be as follows.

\vspace{1em}

[2.87, 2.87, 2.87, 2.87, 2.87, 2.87,...] 

\vspace{1em}

However, if some of the frames vis-a-vis rows are missing 
from events list FITS table, we might get an array as given below.

\vspace{1em}

[2.87, 2.53, 1.01, 2.84, 1.94, 2.87,...]

\vspace{1em}

A false variability can be inferred from the above array. 
To overcome this, the original-time histogram is used. 
The missing frames will be reflected as a reduced number 
of events per bin in original-time histogram. 
For the case above, the original-time histogram would be as follows.

\vspace{1em}

[28.7, 25.3, 10.1, 28.4, 19.4, 28.7,...]

\vspace{1em}

By taking the ratio of two histograms, the real light curve in CPF 
is obtained as follows.

\vspace{1em}

[0.1, 0.1, 0.1, 0.1, 0.1, 0.1,...]

\vspace{1em}

CPF is then converted to CPS by multiplying with the frame-rate 
($\sim$28.7 for 512$\times$512 mode).

\section*{Acknowledgements}

The UVIT project is a result of collaboration between Indian Institute of Astrophysics, Bangalore; the Inter-University Centre for Astronomy and Astrophysics, Pune; Tata Institute of Fundamental Research, Mumbai; many centres of the Indian Space Research Organisation and the Canadian Space Agency.

\balance

\bibliography{references}

\begin{thebibliography}{}
\expandafter\ifx\csname natexlab\endcsname\relax\def\natexlab#1{#1}\fi

\bibitem[{Agrawal(2006)}]{AGRAWAL2006}
Agrawal, P. 2006, Advances in Space Research, 38, 2989 , spectra and Timing of
  Compact X-ray Binaries

\bibitem[{{Astropy Collaboration} {$et~al$.}(2013){Astropy Collaboration},
  {Robitaille}, {Tollerud}, {Greenfield}, {Droettboom}, {Bray}, {Aldcroft},
  {Davis}, {Ginsburg}, {Price-Whelan}, {Kerzendorf}, {Conley}, {Crighton},
  {Barbary}, {Muna}, {Ferguson}, {Grollier}, {Parikh}, {Nair}, {Unther},
  {Deil}, {Woillez}, {Conseil}, {Kramer}, {Turner}, {Singer}, {Fox}, {Weaver},
  {Zabalza}, {Edwards}, {Azalee Bostroem}, {Burke}, {Casey}, {Crawford},
  {Dencheva}, {Ely}, {Jenness}, {Labrie}, {Lim}, {Pierfederici}, {Pontzen},
  {Ptak}, {Refsdal}, {Servillat}, \& {Streicher}}]{astropy:2013}
{Astropy Collaboration}, {Robitaille}, T.~P., {Tollerud}, E.~J., {$et~al$.}
  2013, aap, 558, A33

\bibitem[{{Astropy Collaboration} {$et~al$.}(2018){Astropy Collaboration},
  {Price-Whelan}, {Sip{H{o}}cz}, {G{"u}nther}, {Lim}, {Crawford}, {Conseil},
  {Shupe}, {Craig}, {Dencheva}, {Ginsburg}, {Vand erPlas}, {Bradley},
  {P{'e}rez-Su{'a}rez}, {de Val-Borro}, {Aldcroft}, {Cruz}, {Robitaille},
  {Tollerud}, {Ardelean}, {Babej}, {Bach}, {Bachetti}, {Bakanov}, {Bamford},
  {Barentsen}, {Barmby}, {Baumbach}, {Berry}, {Biscani}, {Boquien}, {Bostroem},
  {Bouma}, {Brammer}, {Bray}, {Breytenbach}, {Buddelmeijer}, {Burke},
  {Calderone}, {Cano Rodr{'i}guez}, {Cara}, {Cardoso}, {Cheedella}, {Copin},
  {Corrales}, {Crichton}, {D'Avella}, {Deil}, {Depagne}, {Dietrich}, {Donath},
  {Droettboom}, {Earl}, {Erben}, {Fabbro}, {Ferreira}, {Finethy}, {Fox},
  {Garrison}, {Gibbons}, {Goldstein}, {Gommers}, {Greco}, {Greenfield},
  {Groener}, {Grollier}, {Hagen}, {Hirst}, {Homeier}, {Horton}, {Hosseinzadeh},
  {Hu}, {Hunkeler}, {Ivezi{'c}}, {Jain}, {Jenness}, {Kanarek}, {Kendrew},
  {Kern}, {Kerzendorf}, {Khvalko}, {King}, {Kirkby}, {Kulkarni}, {Kumar},
  {Lee}, {Lenz}, {Littlefair}, {Ma}, {Macleod}, {Mastropietro}, {McCully},
  {Montagnac}, {Morris}, {Mueller}, {Mumford}, {Muna}, {Murphy}, {Nelson},
  {Nguyen}, {Ninan}, {N{"o}the}, {Ogaz}, {Oh}, {Parejko}, {Parley}, {Pascual},
  {Patil}, {Patil}, {Plunkett}, {Prochaska}, {Rastogi}, {Reddy Janga},
  {Sabater}, {Sakurikar}, {Seifert}, {Sherbert}, {Sherwood-Taylor}, {Shih},
  {Sick}, {Silbiger}, {Singanamalla}, {Singer}, {Sladen}, {Sooley},
  {Sornarajah}, {Streicher}, {Teuben}, {Thomas}, {Tremblay}, {Turner},
  {Terr{'o}n}, {van Kerkwijk}, {de la Vega}, {Watkins}, {Weaver}, {Whitmore},
  {Woillez}, {Zabalza}, \& {Astropy Contributors}}]{astropy:2018}
{Astropy Collaboration}, {Price-Whelan}, A.~M., {Sip{H{o}}cz}, B.~M.,
  {$et~al$.} 2018, aj, 156, 123

\bibitem[{Bradley {$et~al$.}(2020)Bradley, Sipőcz, Robitaille, Tollerud,
  Vinícius, Deil, Barbary, Wilson, Busko, Günther, Cara, Conseil, Bostroem,
  Droettboom, Bray, Bratholm, Lim, Barentsen, Craig, Pascual, Perren, Greco,
  Donath, de~Val-Borro, Kerzendorf, Bach, Weaver, D'Eugenio, Souchereau, \&
  Ferreira}]{photutils}
Bradley, L., Sipőcz, B., Robitaille, T., {$et~al$.} 2020, astropy/photutils:
  1.0.1, doi:10.5281/zenodo.4049061

\bibitem[{Harris {$et~al$.}(2020)Harris, Millman, van~der Walt, Gommers,
  Virtanen, Cournapeau, Wieser, Taylor, Berg, Smith, Kern, Picus, Hoyer, van
  Kerkwijk, Brett, Haldane, Fernández~del Río, Wiebe, Peterson,
  Gérard-Marchant, Sheppard, Reddy, Weckesser, Abbasi, Gohlke, \&
  Oliphant}]{numpy}
Harris, C.~R., Millman, K.~J., van~der Walt, S.~J., {$et~al$.} 2020, Nature,
  585, 357–362

\bibitem[{Hunter(2007)}]{matplotlib}
Hunter, J.~D. 2007, Computing In Science \& Engineering, 9, 90

\bibitem[{Hutchings {$et~al$.}(2007)Hutchings, Postma, Asquin, \&
  Leahy}]{Hutchings_2007}
Hutchings, J.~B., Postma, J., Asquin, D., \& Leahy, D. 2007, Publications of
  the Astronomical Society of the Pacific, 119, 1152

\bibitem[{Maneewongvatana \& Mount(1999)}]{maneewongvatana1999s}
Maneewongvatana, S., \& Mount, D.~M. 1999, in Center for geometric computing
  4th annual workshop on computational geometry, Vol.~2, 1--8

\bibitem[{Postma {$et~al$.}(2011)Postma, Hutchings, \& Leahy}]{Postma_2011}
Postma, J., Hutchings, J.~B., \& Leahy, D. 2011, Publications of the
  Astronomical Society of the Pacific, 123, 833

\bibitem[{Postma \& Leahy(2017)}]{postma2017ccdlab}
Postma, J.~E., \& Leahy, D. 2017, Publications of the Astronomical Society of
  the Pacific, 129, 115002

\bibitem[{Rani {$et~al$.}(2019)Rani, Stalin, \& Goswami}]{rani2019study}
Rani, P., Stalin, C., \& Goswami, K.~D. 2019, Monthly Notices of the Royal
  Astronomical Society, 484, 5113

\bibitem[{Stetson(1987)}]{stetson1987daophot}
Stetson, P.~B. 1987, Publications of the Astronomical Society of the Pacific,
  99, 191

\bibitem[{Tandon {$et~al$.}(2017{\natexlab{a}})Tandon, Stalin, Subramaniam,
  Ghosh, \& Hutchings}]{tandon2017CurrentScience}
Tandon, S., Stalin, C., Subramaniam, A., Ghosh, S., \& Hutchings, J.
  2017{\natexlab{a}}, Current Science (Bangalore), 113, 583

\bibitem[{Tandon {$et~al$.}(2017{\natexlab{b}})Tandon, Subramaniam, Girish,
  Postma, Sankarasubramanian, Sriram, Stalin, Mondal, Sahu, Joseph, Hutchings,
  Ghosh, Barve, George, Kamath, Kathiravan, Kumar, Lancelot, Leahy, Mahesh,
  Mohan, Nagabhushana, Pati, Rao, Sreedhar, \&
  Sreekumar}]{Tandon2017calibration}
Tandon, S.~N., Subramaniam, A., Girish, V., {$et~al$.} 2017{\natexlab{b}}, The
  Astronomical Journal, 154, 128

\bibitem[{Tandon {$et~al$.}(2017{\natexlab{c}})Tandon, Hutchings, Ghosh,
  Subramaniam, Koshy, Girish, Kamath, Kathiravan, Kumar, Lancelot, Mahesh,
  Mohan, Murthy, Nagabhushana, Pati, Postma, Rao, Sankarasubramanian,
  Sreekumar, Sriram, Stalin, Sutaria, Sreedhar, Barve, Mondal, \&
  Sahu}]{Tandon2017performance}
Tandon, S.~N., Hutchings, J.~B., Ghosh, S.~K., {$et~al$.} 2017{\natexlab{c}},
  Journal of Astrophysics and Astronomy, 38, 28

\bibitem[{Tandon {$et~al$.}(2020)Tandon, Postma, Joseph, Devaraj, Subramaniam,
  Barve, George, Ghosh, Girish, Hutchings, Kamath, Kathiravan, Kumar, Lancelot,
  Leahy, Mahesh, Mohan, Nagabhushana, Pati, Rao, Sankarasubramanian, Sriram, \&
  Stalin}]{Tandon_2020}
Tandon, S.~N., Postma, J., Joseph, P., {$et~al$.} 2020, The Astronomical
  Journal, 159, 158

\bibitem[{Virtanen {$et~al$.}(2020)Virtanen, Gommers, Oliphant, Haberland,
  Reddy, Cournapeau, Burovski, Peterson, Weckesser, Bright, {van der Walt},
  Brett, Wilson, Millman, Mayorov, Nelson, Jones, Kern, Larson, Carey, Polat,
  Feng, Moore, {VanderPlas}, Laxalde, Perktold, Cimrman, Henriksen, Quintero,
  Harris, Archibald, Ribeiro, Pedregosa, {van Mulbregt}, \& {SciPy 1.0
  Contributors}}]{scipy}
Virtanen, P., Gommers, R., Oliphant, T.~E., {$et~al$.} 2020, Nature Methods,
  17, 261

\end{thebibliography}

\end{document}